\documentclass{aa}  
\usepackage{graphicx}
\usepackage{natbib}

\bibpunct{(}{)}{;}{a}{}{,}   
\usepackage{txfonts}

\begin{document}
   \title{The $10^5$\,L$_{\odot}$ High-Mass Protostellar Object
     IRAS\,23151+5912}


   \author{H.~Beuther\inst{1}, Q. Zhang\inst{2}, T.R. Hunter\inst{3}, T.K. Sridharan\inst{2}, E.A. Bergin\inst{4}}


   \offprints{H.~Beuther}

   \institute{Max-Planck-Institute for Astronomy, K\"onigstuhl 17, 
              69117 Heidelberg, Germany\\
              \email{beuther@mpia.de}
         \and
              Harvard-Smithsonian Center for Astrophysics, 60 Garden Street,
              Cambridge, MA 02138, USA\\
             \email{name@cfa.harvard.edu}
         \and         
              NRAO, 520 Edgemont Rd,
              Charlottesville, VA 22903\\
              \email{thunter@nrao.edu}
         \and         
             University of Michigan, 825 Dennison Building, 
             500 Church Street, Ann Arbor, MI 48109-1042 \\
             \email{ebergin@umich.edu}
             }
\authorrunning{Beuther et al.}
\titlerunning{SMA observations of IRAS\,23151+5912}

   \date{}

  \abstract
  {While most sources above $10^5$\,L$_{\odot}$ have already formed an
    Ultracompact H{\sc ii} region (UCH{\sc ii}), this is not
    necessarily the case for sources of lower luminosity.
    Characterizing sources in the transition phase, i.e., very
    luminous objects without any detectable free-free emission, is
    important for a general understanding of massive star formation.}
  {Characterizing one of the most luminous High-Mass Protostellar
    Objects (HMPO) that has not yet formed any detectable UCH{\sc ii}
    region.}
  {The region was observed with the Submillimeter Array in three
    different array configurations at 875\,$\mu$m in the submm
    continuum and spectral line emission at sub-arcsecond resolution.}
  {The 875\,$\mu$m submm continuum emission has been resolved into at
    least two condensations. The previously believed driving source of
    one of the outflows, the infrared source IRS1, is $\sim 0.9''$
    offset from the main submm peak. The data do not allow to
    differentiate whether this offset is real, either caused by
    different sources or a shift of the photo-center due to
    scattering, or whether it is only due to poor astrometry of the
    infrared data. Over the entire 4\,GHz bandwidth we detect an
    intermediate dense spectral line forest with 27 lines from 8
    different species, isotopologues or vibrationally-torsionally
    excited states. Temperature estimates based on the CH$_3$OH line
    series result in values of $T(\rm{Peak1})\sim 150\pm 50$\,K and
    $T(\rm{Peak2})\sim 80\pm 30$\,K for the two submm peak positions,
    respectively. The SiO(8--7) red- and blue-shifted line maps
    indicate the presence of two molecular outflows. In contrast, the
    vibrationally-torsionally excited CH$_3$OH line exhibits a
    velocity gradient approximately perpendicular to one of the
    outflows. With a size of approximately 5000\,AU and no Keplerian
    rotation signature, this structure does not resemble a genuine
    accretion disk but rather a larger-scale rotating toroid that may
    harbor a more common accretion disk at its so far unresolved
    center.}
    {}
   

   \maketitle
%

\section{Introduction}
\label{intro}

In contrast to known Ultracompact H{\sc ii} regions (UCH{\sc ii}s)
with luminosities sometimes even exceeding $10^6$\,L$_{\odot}$ (e.g.,
\citealt{wc89}), a close inspection of large samples of younger
High-Mass Protostellar Objects (HMPOs) (e.g.,
\citealt{molinari1996,sridha,faundez2004,beltran2006}) shows that only
very few sources at that evolutionary stage exceed
$10^5$\,L$_{\odot}$. The average luminosity difference between HMPOs
and UCH{\sc ii}s is likely attributed to evolutionary differences,
i.e., the still accreting HMPO will gain more mass eventually being
energetic enough to produce detectable free-free emission at cm
wavelengths and hence evolve into an UCH{\sc ii} region (e.g.,
\citealt{sridha,beuther2006b}).  Observations indicate that below
$10^5$\,L$_{\odot}$ the free-free emission of the forming star is
still potentially quenched, either by the pressure of the infalling
gas, high recombination rates due to high accretion and infall rates,
or by gravitational trapping (e.g., \citealt{walmsley1995,keto2003}).
However, since almost all massive star-forming regions above
$10^5$\,L$_{\odot}$ are detected at cm wavelengths, these processes
appear less feasible in that regime.  Most interferometric
investigations focused either on very luminous UCH{\sc ii} regions
(e.g., \citealt{cesaroni1994,keto2002a,beltran2004,hoare2006}) or on
younger but less luminous HMPOs (e.g.,
\citealt{shepherd1998,zhang1998a,cesaroni1999,beuther2002d,beuther2005c}).
This transition phase of a source that has already a very luminous and
massive accreting protostar, however, that is yet not capable to
ionize enough gas to be detected as a Hypercompact H{\sc ii} region
(HC{\sc ii}), has not been studied well so far. Therefore, here we
present new Submillimeter Array (SMA\footnote{The Submillimeter Array
  is a joint project between the Smithsonian Astrophysical Observatory
  and the Academia Sinica Institute of Astronomy and Astrophysics, and
  is funded by the Smithsonian Institution and the Academia Sinica.})
data observed toward the $10^5$\,L$_{\odot}$ HMPO IRAS\,23151+5912.

The rationale to choose IRAS\,23151+5912 was foremost based on two
criteria: the source had to be very luminous ($\geq
10^5$\,L$_{\odot}$) and it had to be very young, in an evolutionary
stage prior to forming any detectable ultracompact H{\sc ii} region.
The source IRAS\,23151+5912 comprised both features, hence it is one
of the most luminous HMPOs in a pre-UCH{\sc ii} region phase.  The
source at a kinematic distance of 5.7\,kpc is part of a large sample
of 69 candidate HMPOs initially studied by
\citet{sridha,beuther2002a,beuther2002b,beuther2002c}.
IRAS\,23151+5912 shows H$_2$O maser emission and it is not detected in
the cm band at 8\,GHz down to 1\,mJy \citep{tofani1995,sridha}, thus
no significant hypercompact H{\sc ii} region has formed yet
\citep{sridha}. Based on single-dish 1.2\,mm dust continuum
observations the total gas mass within the region is estimated to
$\sim$600\,M$_{\odot}$ \citep{beuther2002a,beuther2002erratum}.
Observations of the CO(2--1) and SiO(2--1) lines with the IRAM\,30\,m
telescope reveal a bipolar outflow with an entrained gas mass of $\sim
20$\,M$_{\odot}$ \citep{beuther2002b}. The derived outflow rate of
$1\times 10^{-3}$\,M$_{\odot}$\,yr$^{-1}$ translates into an estimate
of the accretion rate of the same order of magnitude
\citep{beuther2002b}.  Recent interferometric high-spatial-resolution
SiO(2--1) observations reveal a quasi-parabolic blue-shifted outflow
cone being consistent with a wide-angle wind \citep{qiu2007}.
\citet{weigelt2006} speckle-imaged the region in the near-infrared
K-band, and the nebulosity they find coincides with the blue outflow
wing. Their data rather suggest that the outflow features are produced
by a precessing jet. In addition, \citet{weigelt2006} find an embedded
K-band source IRS1 which they propose to be the driving source of the
molecular outflow.  Furthermore, NH$_3$ inversion lines from the (1,1)
to the (4,4) transitions have been detected with the Effelsberg 100\,m
telescope (unpublished data, Beuther priv. comm.), the excitation
temperature of the (4,4) line is about 200\,K. The NH$_3$ (1,1)/(4,4)
line ratio of only $\sim 2$ indicates warm and dense gas at the core
center. Very recently, \citet{garay2007} detected an unresolved
hypercompact H{\sc ii} region at a sub-mJy flux level toward
IRAS\,23151+5912. Furthermore, they detected about $30''$ apart a more
evolved cometary H{\sc ii} region. While the latter may contribute
part of the total luminoisty of the region ($\sim$20\%,
\citealt{garay2007}), the main power-house of the region very likely
remains the HMPO IRAS\,23151+5912. Therefore, our target source
comprises all features of a luminous High-Mass Protostellar Object at
an early evolutionary stage.  Assuming that the source is still in its
accretion phase, it has not yet reached its final luminosity (e.g.,
\citealt{sridha}), and the total luminosity of the forming cluster
will even exceed $10^5$\,L$_{\odot}$ during its ongoing evolution.

\section{Observations}

We have observed the HMPO IRAS\,23151+5912 with the Submillimeter
Array during three nights between May and November 2005. We used three
different array configurations (compact, extended, very extended) with
unprojected baselines between 16 and 500\,m, resulting at $875\,\mu$m
in a projected baseline range from 8 to 591\,k$\lambda$. The chosen
phase center was R.A. [J2000.0]: $23^h17^m21.^s0$ and Decl. [J2000.0]
$+59^{\circ}28'48.''995$. The velocity of rest is $v_{lsr}\sim
-54.4$\,km\,s$^{-1}$.

For bandpass calibration we used 3C279 and 3C454.3.  The flux scale
was derived in the very extended configuration from observations of
3c279. For the extended and compact configurations, we used 3C454.3
for the relative scaling between the various baselines and then scaled
that absolutely via observations of Uranus.  The flux accuracy is
estimated to be accurate within 20\%. Phase and amplitude calibration
was done via frequent observations of the quasars J0102+584 and BL
Lac, about 13.5$^{\circ}$ and 20.7$^{\circ}$ from the phase center.
The zenith opacity $\tau(\rm{348GHz})$, estimated from the
measurements of the NRAO tipping radiometer located at the Caltech
Submillimeter Observatory, varied between the different observations
nights between $\sim$0.1 and $\sim$0.3 (scaled from the 225\,GHz
measurement, $\tau(348\rm{GHZ})\sim 2.5\times \tau(225\rm{GHz})$). The
receiver operated in a double-sideband mode with an IF band of
4-6\,GHz so that the upper and lower sideband were separated by
10\,GHz. The central frequencies of the upper and lower sideband were
348.2 and 338.2\,GHz, respectively. The correlator had a bandwidth of
2\,GHz and the channel spacing was 0.8125\,MHz.  Measured
double-sideband system temperatures corrected to the top of the
atmosphere were between 100 and 1000\,K, depending on the zenith
opacity and the elevation of the source. Our sensitivity was
dynamic-range limited by the side-lobes of the strongest emission
peaks and thus varied between the line maps of different molecules and
molecular transitions. This limitation was due to the incomplete
sampling of short uv-spacings and the presence of extended structures.
The $1\sigma$ rms for the velocity-integrated molecular line maps (the
velocity ranges for the integrations were chosen for each line
separately depending on the line-widths and intensities) ranged
between 59 and 80\,mJy\,beam$^{-1}$ (Table \ref{rms}). The average
synthesized beam of the spectral line maps was $1'' \times 0.8''$
(Table \ref{rms}).  The 875\,$\mu$m submm continuum image was created
by averaging the line-free parts of both sidebands. The $1\sigma$ rms
of the submm continuum image was $\sim$4.9\,mJy/beam, and the achieved
synthesized beam was $0.57''\times 0.48''$ (P.A.  $86^{\circ}$).  The
initial flagging and calibration was done with the IDL superset MIR
originally developed for the Owens Valley Radio Observatory
\citep{scoville1993} and adapted for the SMA\footnote{The MIR cookbook
  by Charlie Qi can be found at
  http://cfa-www.harvard.edu/$\sim$cqi/mircook.html.}. The imaging and
data analysis were conducted in MIRIAD \citep{sault1995}.

\begin{table}[htb]
  \caption{Peak fluxes densities, rms and synthesized beams of integrated line images in Fig.~\ref{lineimages}.}
\begin{tabular}{lrrr}
\hline \hline
Line & $S_{\rm{peak}}$ & rms & Beam \\
    & $\frac{\rm{mJy}}{\rm{beam}}$  &  $\frac{\rm{mJy}}{\rm{beam}}$ & $''$(P.A.)\\
\hline 
CH$_3$OH$(7_{3,5}-6_{2,4})$ & 1014 & 79 & $0.8\times 0.67$(19) \\
CH$_3$OH$(7_{4,3}-6_{4,3}), v_t=1$ & 617 & 80 & $1.0\times 0.81$(19) \\
C$^{34}$S$(7-6)$ & 411 & 64 & $1.08\times 0.87$(18) \\
SiO$(8-7)$ & 433 & 59 & $1.14\times 0.89$(80) \\
SO$_2(24_{2,22}-23_{3,21})$ & 826 & 60 & $1.07\times 0.84$(80) \\
$^{34}$SO$(8_8-7_7)$ & 977 & 67  & $1.08\times 0.87$(19) \\
\hline \hline
\end{tabular}
\label{rms}
\end{table}

\section{Results}

\subsection{875\,$\mu$m continuum emission in IRAS\,23151+5912}
\label{continuumtext}

\begin{figure}[htb]
\includegraphics[angle=-90,width=8.7cm]{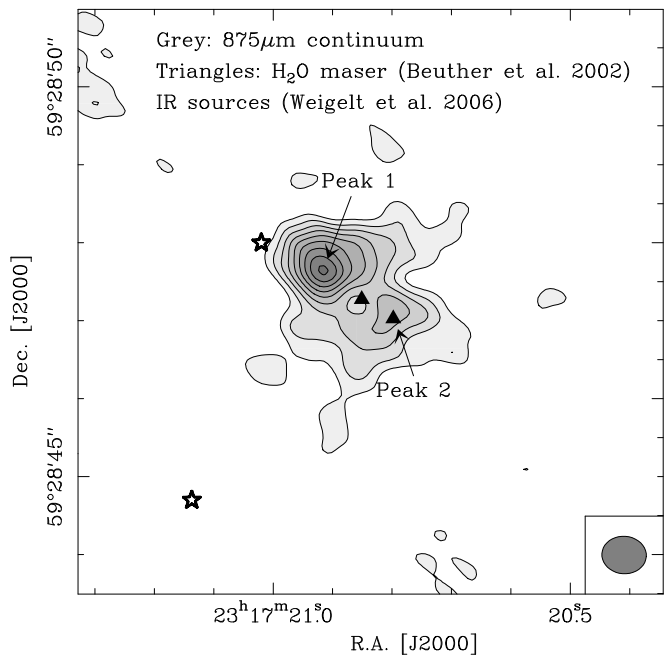}
\caption{The grey-scale with contours shows the 875\,$\mu$m continuum
  emission. The contours go from $\pm$18 ($3\sigma$) to $\pm$98\%
  (step $\pm$10\%) from the peak flux of 80.8\,mJy/beam. The two main
  submm peaks are marked and the triangles and stars show the
  positions of H$_2$O maser and near-infrared peaks from
  \citet{beuther2002c} and \citet{weigelt2006}. The synthesized beam
  of $0.57''\times 0.48''$ is shown at the bottom right.}
\label{continuum}
\end{figure}

Figure \ref{continuum} shows the 875\,$\mu$m continuum image extracted
from the line-free part of the data-cube. The spatial resolution of
$0.57''\times 0.48''$ corresponds to $\sim$3000\,AU at the given
distance of $\sim$5.7\,kpc. The peak and integrated fluxes of the main
peak 1 are about double of those from the minor peak 2 (Table
\ref{submmcont}). However, one of the H$_2$O maser positions is
clearly associated with submm peak 2 and none with submm peak 1. The
second H$_2$O maser position is right between both submm peak
positions. Interestingly, the infrared source IRS1 detected by
\citet{weigelt2006} is offset by about $0.9''$ north-east of the main
submm peak position. The positional accuracy of the interferometric
submm data is estimated to be $\sim 0.1''$. Contrary to that, the
astrometry of the infrared data is based on the 2MASS astrometry, and
this is estimated in this field to be not better than $1''$
(Preibisch, priv. comm.).  Therefore, we cannot unambiguously decide
whether IRS1 and the submm peak 1 are the same or different sources.
Based on the estimated H$_2$ column densities of the order
$10^{24}$\,cm$^{-2}$ (corresponding to visual extinctions of the order
1000, see below), one would not expect to detect near-infrared emission
from the submm peaks.  However, arguments based on the outflow
emission may explain this (see \S\ref{driving}). Furthermore,
\citet{qiu2007} detected more extended 3.4 and 1.3\,mm emission west
of the peak.  Although their 1.3\,mm feature is only at a $2-3\sigma$
level, the morphological similarity with the 3.4\,mm data and the
infrared/SiO outflow stimulates the speculation that this mm emission
may be caused by the outflow. Extrapolating the 1.3\,mm flux density of $\sim
4.25$\,mJy with a typical jet-frequency dependence of $\nu^{0.6}$ to
our continuum frequency of 875\,$\mu$m, the expected submm flux density would
be $S(875\mu\rm{m})\sim 5.4$\,mJy, only barely above our 1$\sigma$
noise level. Therefore, it is no surprise that we cannot detect these
features in the submm data.

\begin{table*}[htb]
\caption{Submm continuum source parameters}
\begin{tabular}{lrrrrrr}
\hline \hline
Source & R.A. & Dec. & $S_{\rm{peak}}$ & $S_{\rm{int}}$ & $M$ & $N$ \\
       & [J2000] & [J2000] & [$\frac{\rm{mJy}}{\rm{beam}}$] & [mJy] & [M$_{\odot}]$ & [10$^{24}$cm$^{-3}$]\\
\hline
Peak 1 & 23:17:20.916 & 59:28:47.65 & 81 & 282 & 26 & 2.3 \\
Peak 2 & 23:17:20.800 & 59:28:47.17 & 44 & 129 & 12 & 1.2 \\
\hline \hline
\end{tabular}
\label{submmcont}
\end{table*}

Table \ref{submmcont} lists the absolute source positions, their
875\,$\mu$m peak fluxes and the integrated flux density approximately
associated with the two sub-sources. Assuming that the observed submm
continuum emission is due to optically thin dust emission, we can
calculate gas masses and gas column densities following
\citet{hildebrand1983} and \citet{beuther2002a,beuther2002erratum}.
Since IRAS23151+5912 is in a very early evolutionary stage we use the
dust temperature derived from the IRAS data (68\,K, \citealt{sridha})
and a dust opacity index $\beta = 2$, corresponding to a dust opacity
per unit mass of $\kappa(875\mu\rm{m}) \sim 0.8$\,cm$^2$g$^{-1}$.
Given the uncertainties in $\beta$ and $T$, we estimate the masses to
be correct within a factor 2-5.  Table \ref{submmcont} gives the
derived masses and column densities separated for each of the two
peaks. While the two core masses vary between 26 and 12\,M$_{\sun}$,
the column densities are of the order $10^{24}$\,cm$^{-2}$,
corresponding to extremely high visual extinctions of the order 1000
($A_v=N_{\rm{H}}/0.94\times 10^{21}$, \citealt{frerking1982}).
Integrating the whole submm continuum emission shown in
Fig.~\ref{continuum} ($S_{\rm{int}} \sim$404\,mJy), the total detected
mass is $\sim$38\,M$_{\odot}$. Comparing the integrated flux density
with the peak flux observed with the SCUBA array on the JCMT ($\sim
15''$ resolution) at 850\,$\mu$m of 1890\,mJy \citep{williams2004},
approximately 80\% of the total flux is filtered out in our
interferometer observations\footnote{Using only the compact
  configuration data, still approximately 60\% of the single-dish
  emission remain filtered out.}. The 3$\sigma$ rms of the continuum
image of $\sim$14.7\,mJy corresponds to a mass sensitivity of
$\sim$1.4\,M$_{\odot}$ at 68\,K. Assuming lower temperatures offset
from the peak, the 3$\sigma$ value corresponds to a higher mass
sensitivity, e.g., $\sim$6.4\,M$_{\odot}$ at 20\,K

\subsection{Submm spectral lines in IRAS\,23151+5912}
\label{spectrallines}

\begin{figure}[htb]
\includegraphics[angle=-90,width=8.7cm]{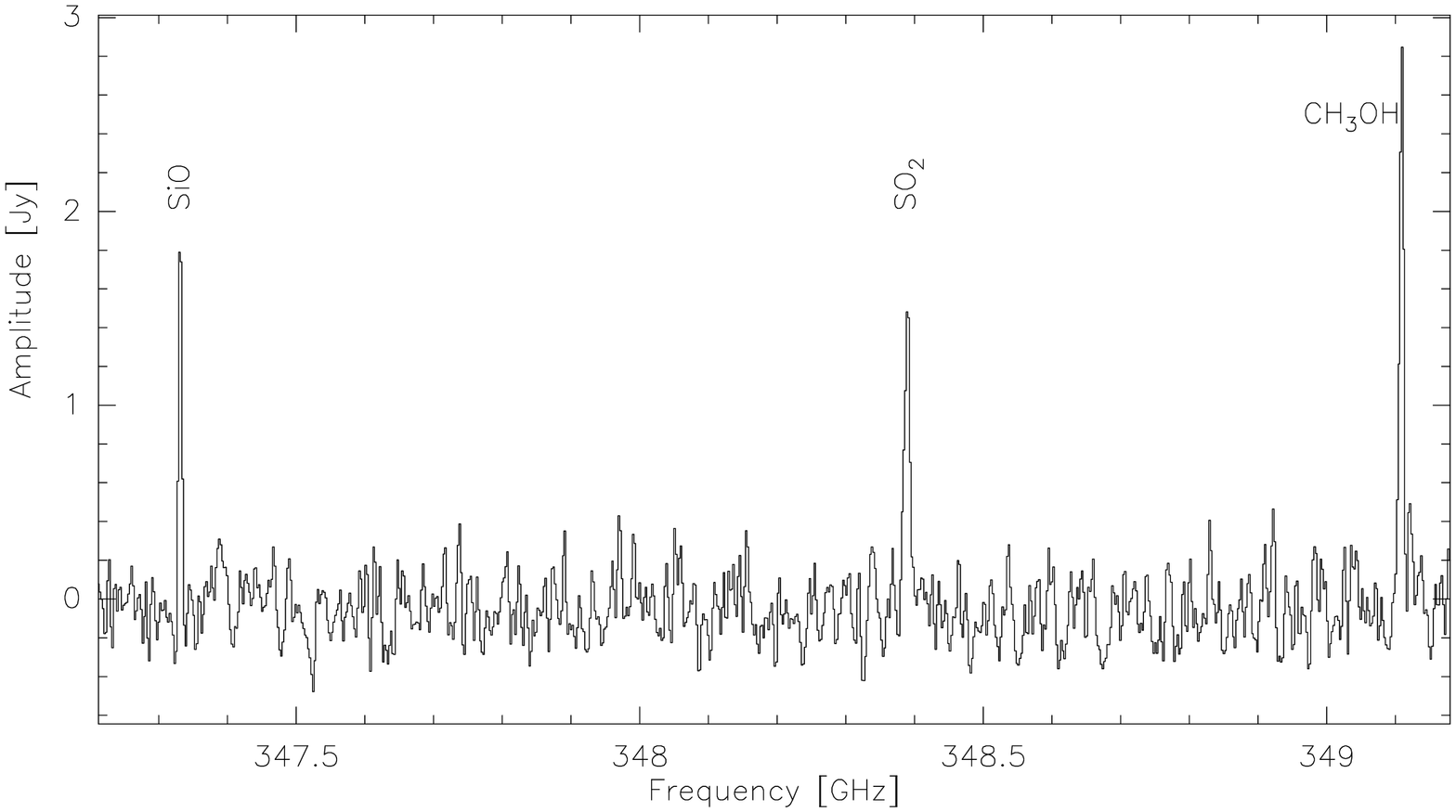}\\
\includegraphics[angle=-90,width=8.7cm]{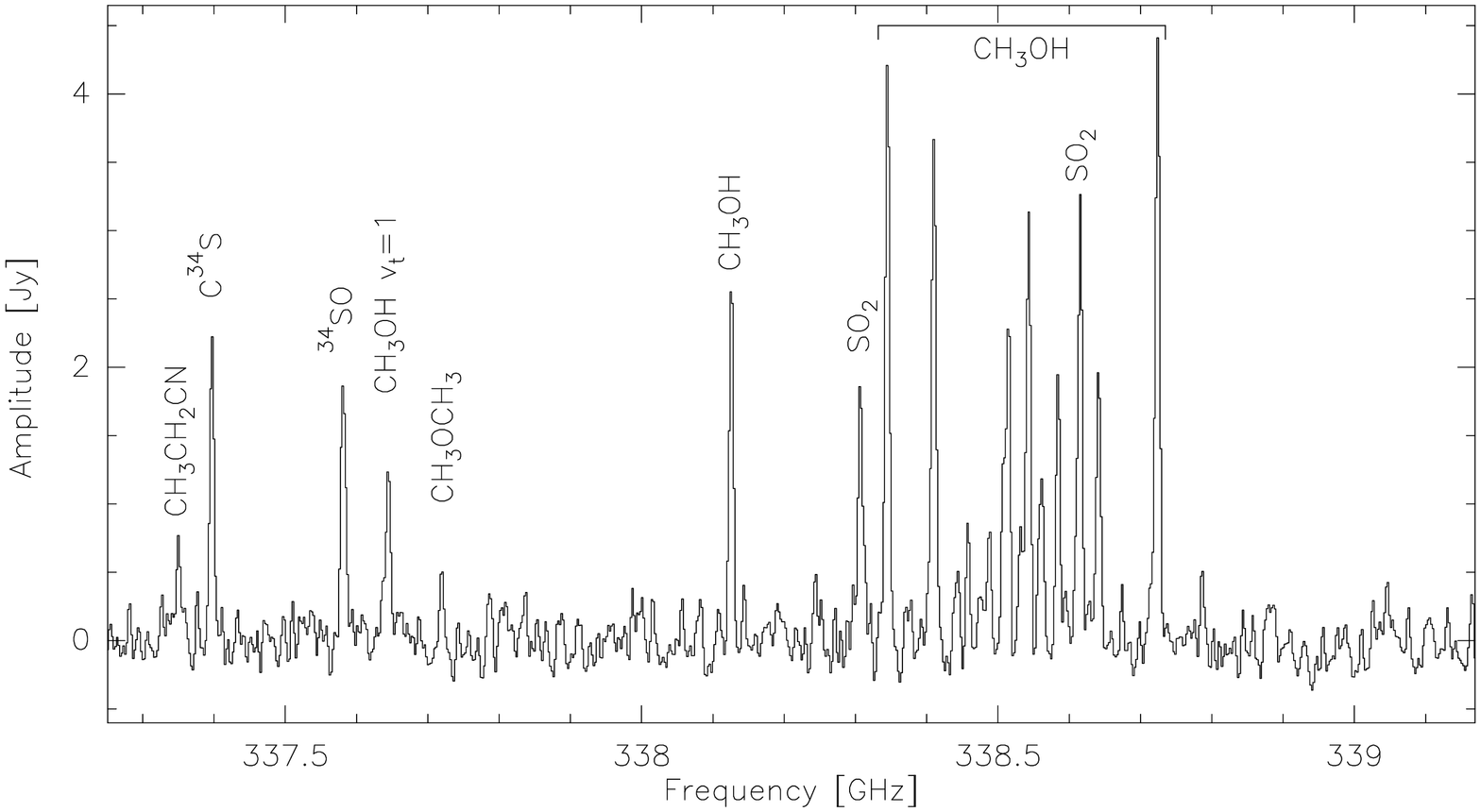}
\caption{Upper and lower sideband (top and bottom panel) spectra taken
  in the uv-domain on a short baseline of $\sim$14\,m. The chosen
  spectral resolution is 2\,km/s.}
\label{uvspectra}
\end{figure}

The spectral bandwidths of 4\,GHz in total revealed 27 spectral lines
(Fig.\ref{uvspectra} and Table \ref{linelist}). We detected a series
of ground-state and vibrationally-torsionally excited CH$_3$OH lines,
some sulphur-bearing species ($^{34}$SO, SO$_2$, C$^{34}$S), the
outflow- and shock-tracing SiO and a few more complex molecules like
CH$_3$CH$_2$CN and CH$_3$OCH$_3$. The excitation temperatures of
especially the CH$_3$OH lines cover a broad range from approximately
40 to 360\,K (Table \ref{linelist}).

\begin{table}[htb]
\caption{Line parameters}
\begin{tabular}{lrrr}
\hline
\hline
Freq. & Line & $E_u$ & $1\sigma$ \\
GHz &      & K & [$\frac{\rm{mJy}}{\rm{beam}}$] \\
\hline
337.348 & CH$_3$CH$_2$CN$(38_{3,36}-37_{3,35})$        & 328 \\
337.397 & C$^{34}$S(7--6)                              & 65 & 60 \\
337.582 & $^{34}$SO$(8_8-7_7)$                         & 86 & 70 \\
337.642 & CH$_3$OH$(7_{1,7}-6_{1,6})$E($v_t$=1)        & 356& 80  \\
337.722 & CH$_3$OCH$_3(7_{4,4}-6_{3,3})$EE             & 48 \\
338.125 & CH$_3$OH$(7_{0,7}-6_{0,6})$E                 & 78 \\
338.306 & SO$_2$$(14_{4,14}-18_{3,15})$                & 197 \\
338.345 & CH$_3$OH$(7_{1,7}-6_{1,6})$E                 & 71 \\
338.409 & CH$_3$OH$(7_{0,7}-6_{0,6})$A                 & 65 \\
338.457 & CH$_3$OH$(7_{5,2}-6_{5,1})$E                 & 189 \\
338.486 & CH$_3$OH$(7_{5,3}-6_{5,2})$A                 & 203 \\
        & CH$_3$OH$(7_{5,2}-6_{5,1})$A$^-$             & 203 \\
338.513 & CH$_3$OH$(7_{4,4}-6_{4,3})$A$^-$             & 145 \\
        & CH$_3$OH$(7_{4,3}-6_{4,2})$A                 & 145 \\
        & CH$_3$OH$(7_{2,6}-6_{2,5})$A$^-$             & 103 \\
338.530 & CH$_3$OH$(7_{4,3}-6_{4,2})$E                 & 161 \\
338.541 & CH$_3$OH$(7_{3,5}-6_{3,4})$A$^+$             & 115 \\
338.543 & CH$_3$OH$(7_{3,4}-6_{3,3})$A$^-$             & 115 \\
338.560 & CH$_3$OH$(7_{3,5}-6_{3,4})$E                 & 128 \\
338.583 & CH$_3$OH$(7_{3,4}-6_{3,3})$E                 & 113 \\
338.612 & SO$_2(20_{1,19}-19_{2,18})$                  & 199 \\
338.640 & CH$_3$OH$(7_{2,5}-6_{2,4})$A                 & 103 \\
338.722 & CH$_3$OH$(7_{2,5}-6_{2,4})$E                 & 87 & 80 \\
338.723 & CH$_3$OH$(7_{2,6}-6_{2,5})$E                 & 91 \\
347.331 & $^{28}$SiO(8--7)                             & 75 & 60 \\
348.388 & SO$_2(24_{2,22}-23_{3,21})$                 & 293 & 58 \\
349.107 & CH$_3$OH$(14_{1,13}-14_{0,14})$              & 43 \\
\hline
\hline
\end{tabular}
\label{linelist}
\end{table}

\begin{figure*}[htb]
\includegraphics[angle=-90,width=18cm]{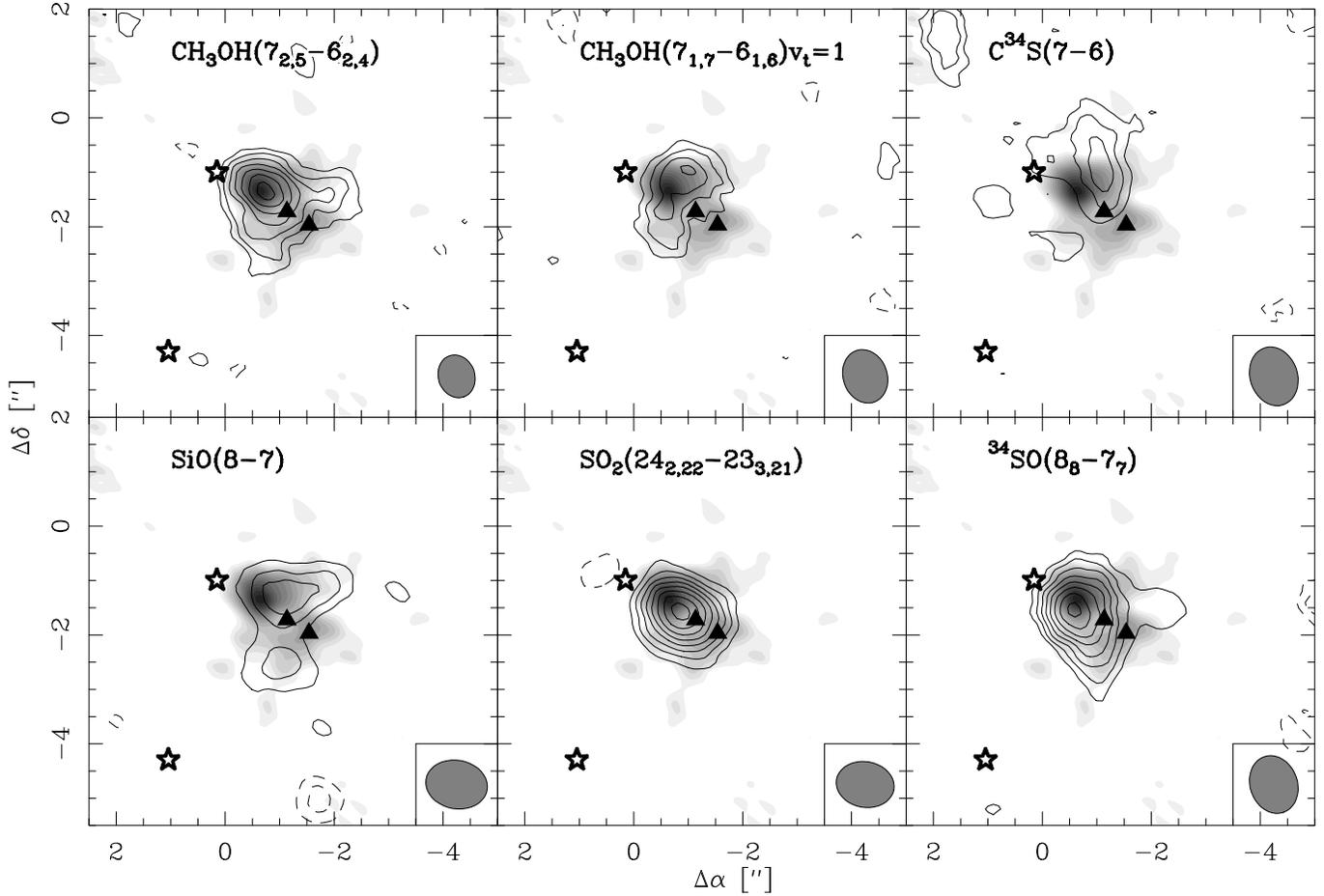}
\caption{Integrated images in IRAS\,23151+5912 of the molecules
  identified in Figure \ref{uvspectra} that are strong enough for
  imaging. The species are labeld in each panel and the synthesized
  beams are shown at the bottom left of each panel. The grey-scale
  shows the submm continuum emission as presented in
  Fig.~\ref{continuum}. Triangles and stars mark the H$_2$O maser and
  near-infrared peaks from \citet{beuther2002c} and
  \citet{weigelt2006}. The contour levels start at $\pm 3\sigma$ and
  continue in $\pm 1.5\sigma$ steps, the $1\sigma$ values are given in
  Table \ref{linelist}.}
\label{lineimages}
\end{figure*}

Excluding the two spectral lines from the more complex molecules
CH$_3$CH$_2$CN and CH$_3$OCH$_3$ that were too weak for imaging, we
were able to image the molecular emission of all other species
(Fig.\ref{lineimages}). The three spectral line images from CH$_3$OH,
SO$_2$ and $^{34}$SO all peak in the near vicinity of the main submm
peak and show emission approximately associated with the region of
submm continuum emission.  This is partly different for the three
other line images. The integrated SiO(8--7) emission shows a
relatively complex morphology which is largely caused by the molecular
outflows in this region (see \ref{outflows}). The
virbrationally-torsionally excited CH$_3$OH emission shows a
double-peaked integrated map around the main submm peak 1 and may be
attributed to rotation from a larger-scale toroid around the most
massive protostellar obect(s) (see \ref{disk}).  Furthermore, the
C$^{34}$S emission is centered offset from the submm continuum peaks
and shows relatively strong more extended emission reminiscent to the
recent observations of the hot molecular core G29.96
\citep{beuther2007d}.

Compared with hot cores like Orion-KL \citep{beuther2005a} or G29.96
\citep{beuther2007d} IRAS\,23151+5912 exhibits relatively few spectral
lines, however, it shows more lines than lower luminosity HMPOs like
IRAS\,05358+3543 (Leurini et al. subm. to A\&A). In a follow-up paper we
present a more detailed comparison between all these sources with
a special emphasis on the chemical diversity and a potential chemical
evolutionary sequence (Beuther et al. subm. to A\&A).

\section{Discussion}

\subsection{Multiple molecular outflows}
\label{outflows}

\begin{figure}[htb]
\includegraphics[angle=-90,width=8.7cm]{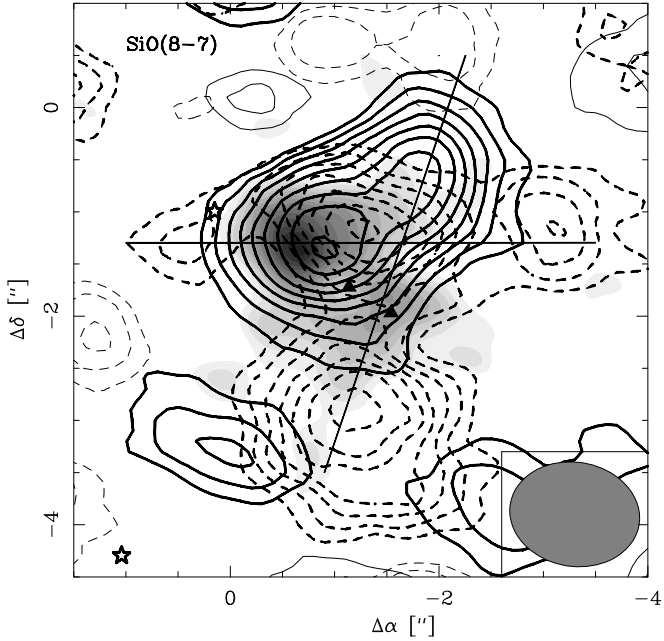}
\caption{The full and dashed contours show the blue- and red-shifted
  SiO(8--7) emission for IRAS\,23151+5912 integrated from $-60$ to
  $-56$ and from $-54$ to $-51$\,km/s, respectively. Thick and thin
  lines present positive and negative features, the latter ones being
  due to the incomplete uv-coverage of interferometer datasets.  The
  two lines highlight the two tentatively identified molecular
  outflows, and the synthesized beam is shown at the bottom-right.
  The grey-scale shows the submm continuum emission as presented in
  Fig.~\ref{continuum}. Triangles and stars mark the H$_2$O maser and
  near-infrared peaks from \citet{beuther2002c} and
  \citet{weigelt2006}. The contour levels go from 18 to 98\% (step
  10\%) of the peak emission with blue- and red-shifted peak values of
  552 and 445\,mJy\,beam$^{-1}$, respectively.}
\label{sio}
\end{figure}

The SiO(8--7) emission allows us to investigate the molecular
outflow(s) in this region. However, in comparison to the SiO(2--1)
emission with a velocity spread from $-$70 to $-$48\,km\,s$^{-1}$
\citep{qiu2007}, the observable total velocity spread in the $J=8-7$
line is only from $-$60 to $-$51\,km\,s$^{-1}$, relatively close to
the velocity of the ambient gas of $-54.4$\,km\,s$^{-1}$
\citep{sridha}.  The difference between the $J=2-1$ and $J=8-7$ lines
can likely be explained by the higher excitation temperature of the
latter line. The energy levels above ground $E_u/k$ are 6 and 75\,K
for the lower and higher excited lines, respectively.  With typical
gas temperatures in molecular outflows of the order 30\,K (e.g.,
\citealt{cabrit1992}), the $J-8-7$ line is much harder to excite and
hence more difficult to detect at all. Separating spectral core from
line wing emission is not an easy task when no pronounced broad line
wings are observable, however, since SiO is believe to be produced
almost exclusively by sputtering from dust grains (e.g.,
\citealt{schilke1997a}), it is likely that most of the SiO emission is
caused by the molcular outflow. Therefore, one can produce outflow
maps from SiO data going closer to the ambient gas velocities than it
is feasible for more common molecules like CO.  Hence, we imaged the
blue- and red-shifted SiO(8--7) emission in the intervals
[-60,-56]\,km/s (blue) and [-54,-51]\,km/s (red), respectively. Figure
\ref{sio} presents this SiO line wing maps as overlay on the submm
continuum emission.  The blue- and red-shifted emission are spatially
distinct, but the morphology is more ambiguous than would be expected
from the previous mm and near-infrared outflow observations (e.g.,
\citealt{beuther2002b,weigelt2006,qiu2007}). These previous
observations clearly identify one outflow in approximately east-west
direction emanating from the region containing the main submm peak and
the infrared source IRS1. This east-west outflow can be recognized in
the new SiO(8--7) data as well, although one should note that the
geometric center of the blue- and red-shifted peak is not exactly the
main submm continuum peak but offset about $0.5''$ to the west.
Furthermore, there is considerable additional SiO blue and red-shifted
emission north and south of the continuum sources, respectively.  This
emission is offset from the dust continuum peaks and hence unlikley be
caused by the ambient gas. Since we have at least two embedded
protostellar sources, we suggest that we are witnessing two outflows
emanating from the region as well.  Reinspecting the SiO(2--1)
observations from \citet{qiu2007}, one finds that their lower
excitation data show an additional red-shifted isolated peak toward
the south as well. Our SMA data were observed in three different
configurations having a relatively good uv-sampling and hence spatial
coverage, however, the southern peak from \citet{qiu2007} is close to
the edge of our primary beam and thus difficult to detect anyway. It
is likely that the SiO observations by \citet{qiu2007} filtered out
more of the extended emission which could cause the relative isolation
of the southern SiO peak in their data.

\subsection{The outflow driving sources}
\label{driving}

Based on a faint jet-like infrared feature emanating from IRS1,
\citet{weigelt2006} propose that this should be the driving source of
the east-west molecular outflow. In contrast to that, one would expect
that powerful outflows like that in IRAS\,23151+5912 are driven by
younger and more deeply embedded sources like the submm peak 1. While
we cannot astrometrically decide whether the two sources are the same
or not, the outflow associations indicate that the infrared and submm
emission may indeed emanate from the same massive protostellar object.
Furthermore, due to scattering effects within embedded objects it is
possible that the emission peak shifts at near-infrared wavelengths in
the direction of the steepest density gradient which is toward the
outflow cavity (e.g., \citealt{eisner2005,stark2006}). Therefore,
while an outflow cavity allows to penetrate deeper into the core than
possible in spherical symmetry, the scattering effects could cause
even a real offset between the infrared and submm source. Hence,
although the near-infared source IRS1 and the submm peak 1 may be two
separate sources, the data are also consistent with both emission
peaks potentially being due to the same embedded young and massive
protostellar source that could drive the east-west outflow.

\begin{figure}[htb]
\includegraphics[angle=-90,width=8.7cm]{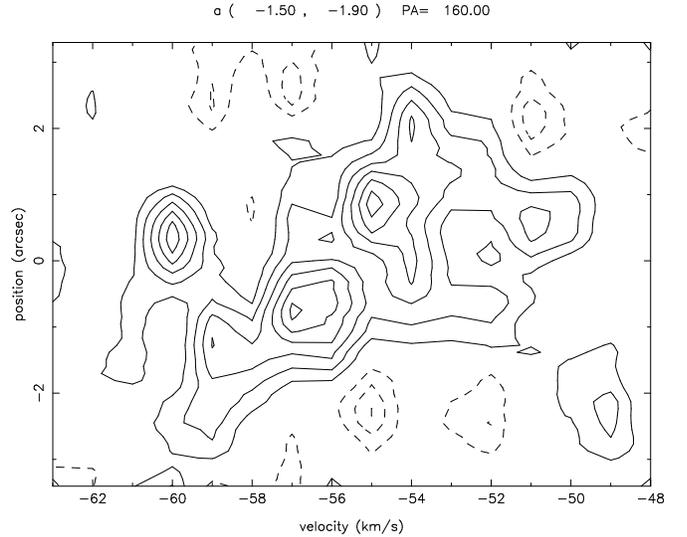}
\caption{SiO(8-7) position-velocity digram of the north-east south-west outflow
  centered on submm peak 2 (offset $-1.5''/-1.9''$ from the phase
  center) with a position angle of 160 degrees from north. Negative
  offsets are going north-west. The contour levels go from $\pm 0.122$
  to $\pm0.732$\,Jy\,beam$^{-1}$ in steps of $\pm
  0.122$\,Jy\,beam$^{-1}$.}
\label{pv_sio}
\end{figure}

The association of one of the maser peaks with submm peak 2
potentially indicates that this source could be driving a molecular
outflow as well. To check this hypothesis, we produced a
position-velocity diagram through submm peak 2 at a position angle of
160 degrees from north, proximately the orientation of the 2nd
proposed outflow (Fig.~\ref{pv_sio}). While it is less clear for the
blue-shifted emission, the red-shifted side shows a clear increase of
velocity with distance from submm peak 2, resembling the Hubble-law
for molecular outflows (e.g., \citealt{richer2000}).  Based on the
morphology and the position-velocity structure of the north-east
south-west outflow, we suggest that this outflow is driven by the
embedded source within submm peak 2.

The second H$_2$O maser position right between the two submm peaks may
either be caused solely by shock interactions of a molecular outflow
with the ambient gas, or it could be associated with a low-mass source
which is below our detection threshhold (3$\sigma$ continuum
sensitivity corresponds to $\sim$1.4\,M$_{\odot}$ at 68\,K or a higher
value of $\sim$6.4\,M$_{\odot}$ at lower temperatures of 20\,K,
\S\ref{continuumtext}) and hence not possible to be identified as an
additional weak submm peak. Although H$_2$O masers are more often
found toward massive star-forming regions (e.g.,
\citealt{forster1999}), there exist several examples of H$_2$O maser
emission from low-mass regions \citep{wilking1994,claussen1998} as
well as from lower-mass companions in regions of high-mass star
formation \citep{hunter1999}.

\subsection{The CH$_3$OH line forest}

\begin{figure}[htb]
\includegraphics[angle=-90,width=8.7cm]{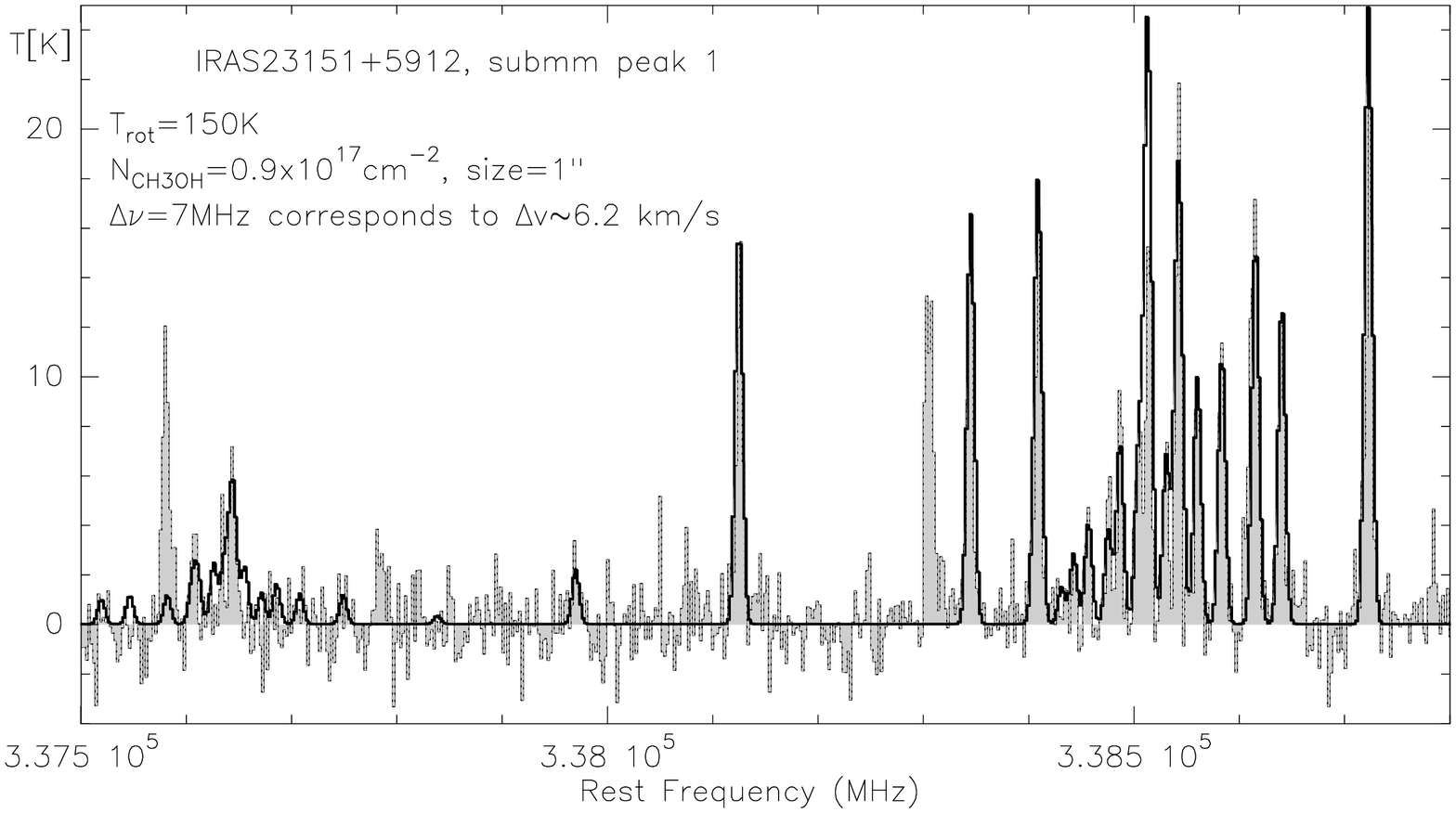}\\
\includegraphics[angle=-90,width=8.7cm]{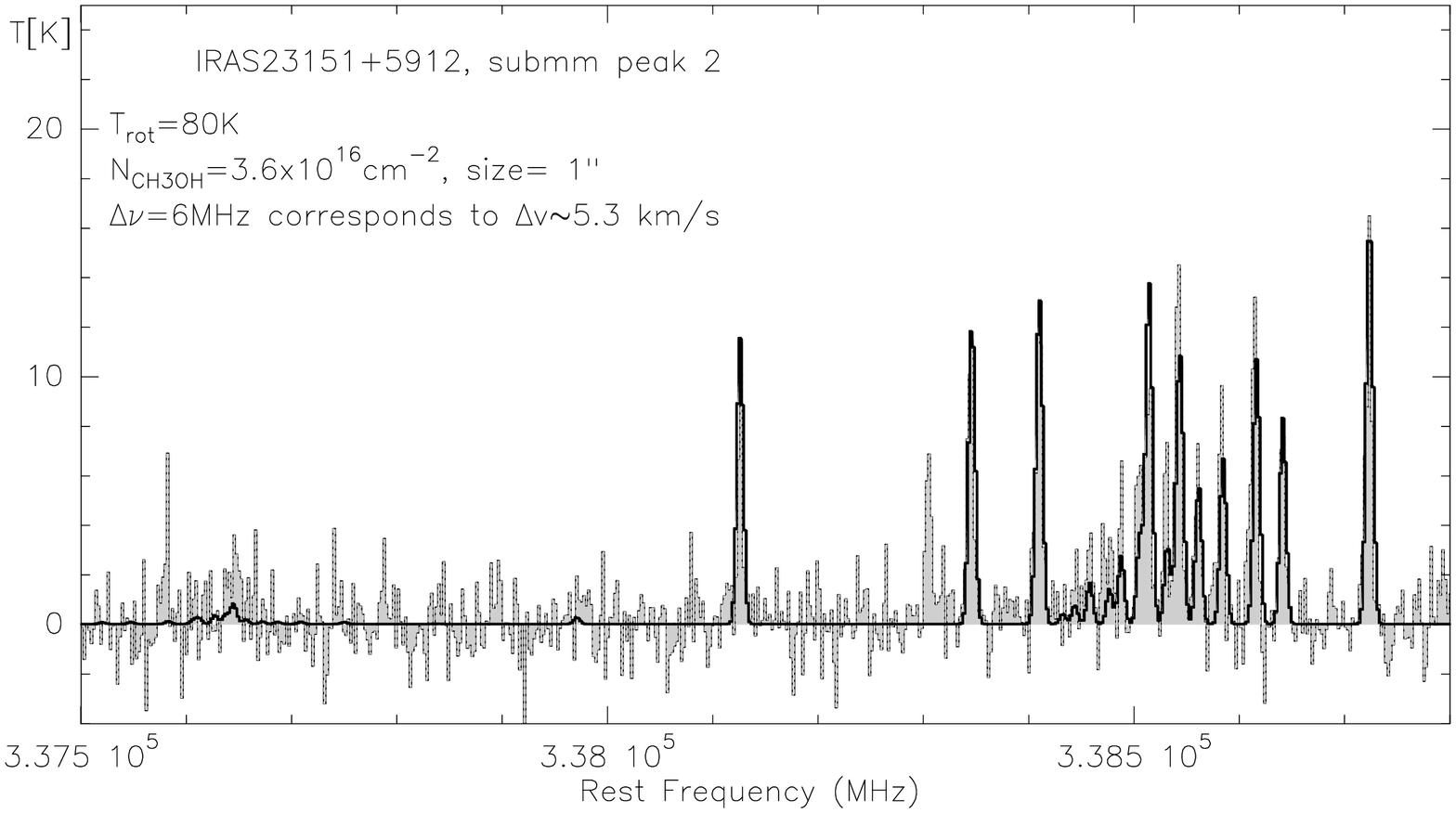}
\caption{Lower sideband spectra toward the two submm peaks. The
  grey-scale and dotted contours show the orginal spectrum at 1.1" x
  0.9" resolution, and the thick solid lines present synthetic spectra
  assuming Local Thermodynamic equilibrium with the parameters stated
  in the figure.}
\label{ch3ohfit}
\end{figure}

The detection of 19 CH$_3$OH lines with excitation temperatures
between 43 and 356\,K allows the temperature to be estimated toward
the two submm continuum peaks. We modeled the whole lower sideband
CH$_3$OH spectrum toward both position in the local thermodynamic
equilibrium approximation using the XCLASS superset to the CLASS
software developed by Peter Schilke (priv.~comm.). This software
package uses the line catalogs from JPL and CDMS
\citep{poynter1985,mueller2001}. The important source dependent model
input parameters are the temperature, the column density, the source
size and the line-width. While the latter one is relatively easy to
determine from the data, the former three parameters are partly
degenerate and various permutations give reasonable fits. Figure
\ref{ch3ohfit} shows the observed lower sideband spectra with
reasonable model spectra overlayed, the exact model input parameters
are stated in each panel. The most important value for the
characterization of the two cores is the temperature.  Therefore, we
went through the parameter space of mainly temperatures and column
density to estimate the uncertainty. The derived temperatures with
approximate error-bars for the two positions are $T(\rm{Peak1})\sim
150\pm 50$\,K and $T(\rm{Peak2})\sim 80\pm 30$\,K. Although the
error-bars are large, as expected from the detection of the
vibrationally-torsionally excited line at the location of submm peak
1, the data indicate that temperatures toward that peak are likely to
be higher.

\subsection{Rotational signatures}
\label{disk}

\begin{figure}[htb]
\includegraphics[angle=-90,width=7.2cm]{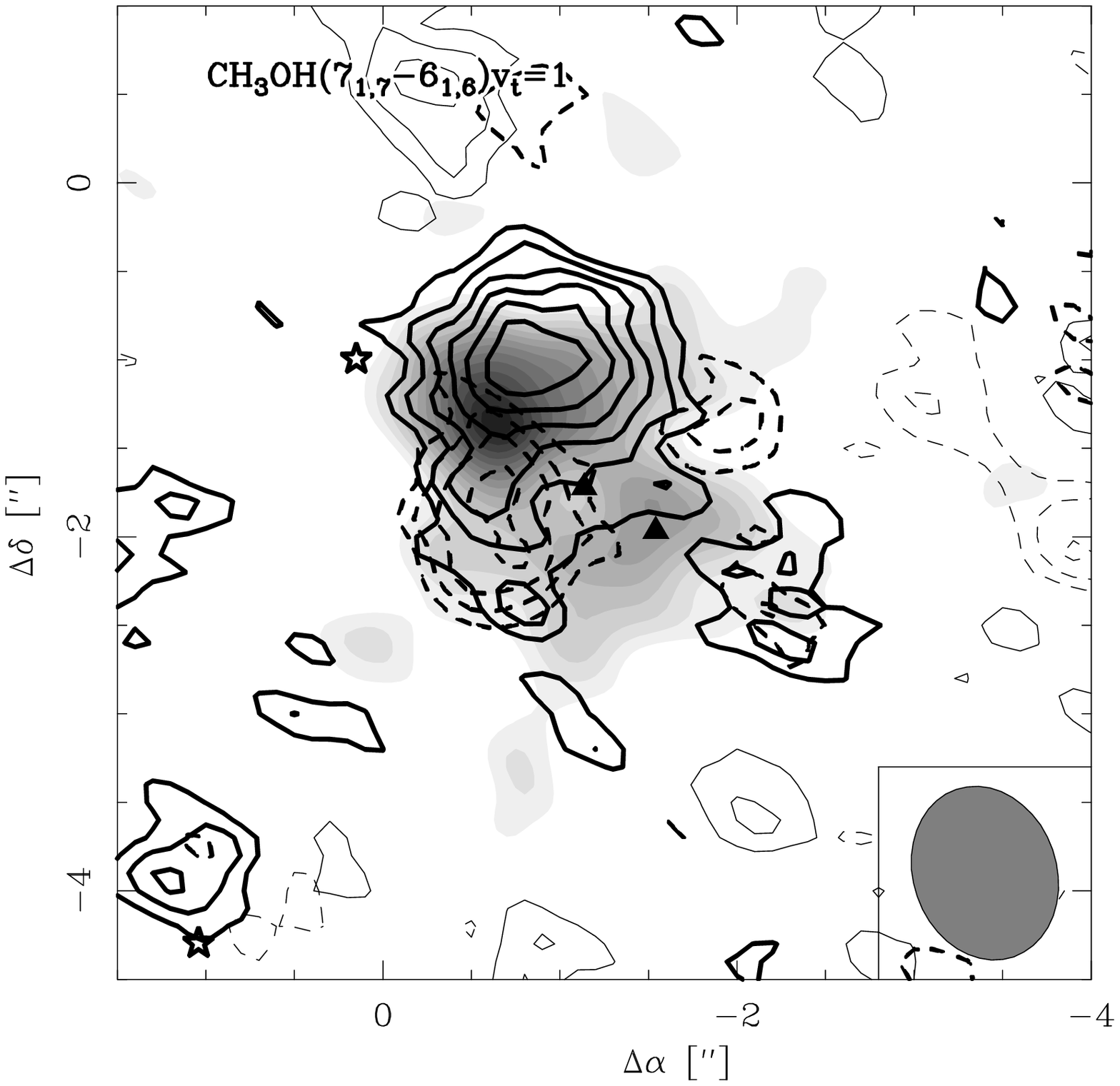}\\
\includegraphics[angle=-90,width=7.2cm]{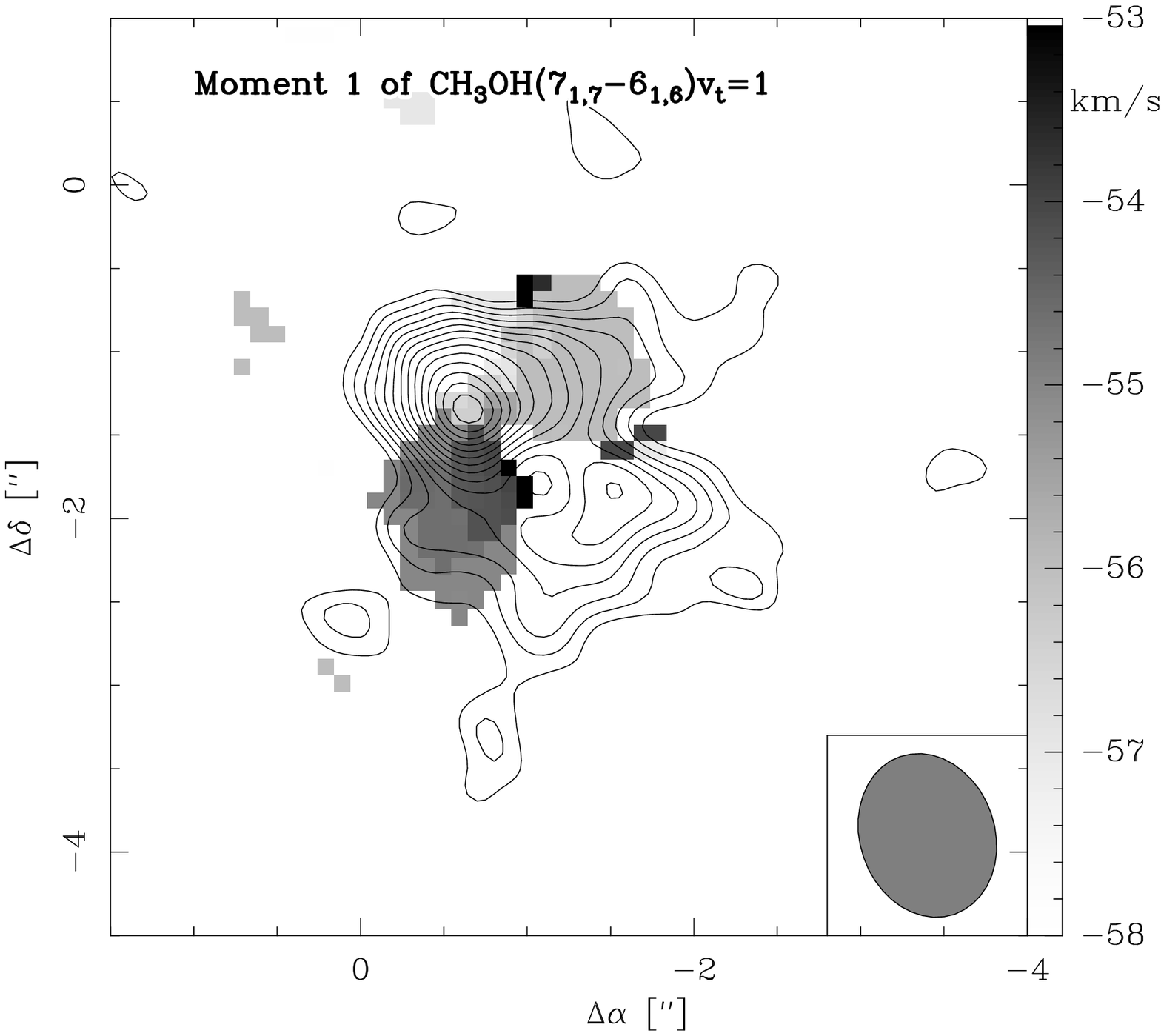}\\
\includegraphics[angle=-90,width=7.2cm]{pv_ch3oh_vt1.ps}
\caption{CH$_3$OH$(7_{1,7}-6_{1,6})v_t=1$ emission for
  IRAS\,23151+5912. {\it Top-panel:} The full/dashed contours show the
  blue-/red-shifted emission integrated from $-58.5$ to $-55.5$ and
  from $-54$ to $-52$\,km/s, respectively. Thick and thin lines
  present positive and negative features, the latter ones being due to
  the incomplete uv-coverage of interferometer datasets. The
  grey-scale shows the submm continuum emission as presented in
  Fig.~\ref{continuum}. Triangles and stars mark the H$_2$O maser and
  near-infrared peaks \citep{beuther2002c,weigelt2006}.  The contour
  levels start at the $\pm 2\sigma$ levels and continue in
  $\pm1\sigma$ steps with the $1\sigma$ levels of 87 and
  100\,mJy\,beam$^{-1}$ for the blue- and red-shifted maps,
  respectively. {\it Middle panel:} The grey-scale shows the 1st
  moment map and the contours outline the submm continuum emission as
  presented in Fig.~\ref{continuum} {\it (The online-version shows
    this panel in color.)}. {\it Bottom panel:} Position-velocity
  diagram through the submm peak position (offset $-0.6''/-1.3''$ from
  the phase reference center) in north-south direction. The contour
  levels go from $\pm 0.136$ to $\pm0.817$\,Jy\,beam$^{-1}$ in steps
  of $\pm 0.136$\,Jy\,beam$^{-1}$.}
\label{ch3oh_vt1}
\end{figure}

One of the missing links in understanding high-mass star formation is
still the detailed understanding of massive accretion disks (e.g.,
\citealt{cesaroni2006,beuther2006b}). Therefore, we searched the
spectral-line data-cube for rotational signatures perpendicular to any
of the molecular outflows. Similar to some other recently investigated
regions (e.g., IRAS\,18089-1732 \citealt{beuther2005c}, G29.96
\citealt{beuther2007d}), most of the lines do not show any coherent
velocity signature indicative of rotation. The only exception is the
vibrationally-torsionally excited methanol line
CH$_3$OH$(7_{1,7}-6_{1,6})v_t=1$ which shows a blue-red velocity shift
in north-south direction around submm peak 1 (Fig.~\ref{ch3oh_vt1}).
Since CH$_3$OH $v_t=1$ has an upper level excitation temperature of
$\sim$356\,K (Table \ref{linelist}) it traces only the warm gas likely
excited by the central growing massive protostar. Furthermore, this
velocity gradient is perpendicular to the main outflow
(\S\ref{outflows} \& Fig.~\ref{sio}), which likely indicates
rotational signatures caused by an underlying collapsing rotational
core-disk system (e.g., \citealt{cesaroni2005b}). The separation
between the two blue- and red-shifted CH$_3$OH $v_t=1$ peaks is
approximately 4300\,AU. In addition to the relatively large size of
the rotating structure, the moment 1 map and position-velocity digram
in Figure \ref{ch3oh_vt1} do not show a Keplerian velocity pattern as
observed for many low-mass sources (e.g., \citealt{simon2000}). A
Keplerian profile would show the largest velocities close to the
center with a velocity decrease proportional to $r^{-1/2}$, whereas
here we see even a velocity increase with distance from the center.
Therefore, this structure is not a Keplerian accretion disk but it
rather resembles a larger-scale rotating toroid that may harbor a
genuine accretion disk at its so far unresolved center (e.g.,
\citealt{cesaroni2005b}).

Although this structure is apparently not in Keplerian rotation, one
can nevertheless estimate an approximate binding mass of this rotating
structure $M_{\rm{rot}}$ assuming equilibrium between the rotational
and gravitational forces at the outer radius. Then one gets

\begin{eqnarray}
M_{\rm{rot}} & = & \frac{v^2r}{G} \label{eq1} \\
\Rightarrow M_{\rm{rot}}[\rm{M_{\odot}}] & = &  1.13\,10^{-3} \times v^2[\rm{km/s}]\times r[\rm{AU}] \label{eq2}
\end{eqnarray}

\noindent where $r$ is the radius of the rotating structure
($\sim$2150\,AU, half the peak separation in Fig.\ref{ch3oh_vt1} top
panel), and $v$ the Half Width Zero Intensity (HWZI) of the spectral
line ($\sim$3\,km\,s$^{-1}$, half the velocity regime used for the
integration in Fig.\ref{ch3oh_vt1} top panel).  Equations \ref{eq1} \&
\ref{eq2} have to be divided by sin$^2(i)$ where $i$ is the unknown
inclination angle between the plane of the rotating structure and the
plane of the sky ($i=90^{\circ}$ for an edge-on system). With the
given values we get a mass $M_{\rm{rot}}$ of $\sim
22$/(sin$^2(i)$)\,M$_{\odot}$, similar to the mass value derived for
Peak 1 from the submm continuum emission (Table \ref{submmcont}).
Assuming an inclination angle of 45 degrees, the rotationally
supported mass would be even 44\,M$_{\odot}$. Although higher than the
gas mass derived from the dust continuum emission, this is still
within the error margins of the dust continuum calculation which is
usually estimated to be correct within factors 2-5
\citep{beuther2002a}. Assuming that the submm continuum emission
largely stems from the disk and/or rotating envelope, this implies
that the disk/envelope mass has to be of the same order than the mass
of the central embedded protostar.  This is different to typical
low-mass protostar-disk systems where the disk-mass is usually
negligable to the protostellar mass. However, one has to note that
such typical low-mass systems are usually more evolved and have
already dispersed much of their envelopes whereas IRAS\,23151+5912 is
still deeply embedded within its natal core. In addition to this, as
shown above, the rotating structure is not in real Kelplerian
rotation, and therefore the assumption of equilibrium between
gravitational and rotational force does exactly hold from the
beginning.

\section{Conclusion and Summary}

We present submm line and continuum observations obtained at
sub-arcsecond spatial resolution with the SMA toward the very young
$10^5$\,L$_{\odot}$ High-Mass Protostellar Object IRAS\,23151+5912.
The submm continuum emission is resolved into two sub-sources, each of
them likely driving one of the molecular outflows observed in
SiO(8-7). The main submm continuum peak is approximately $1''$ offset
from a recently identified near-infrared source which is suggested to
drive one of the molecular outflows \citep{weigelt2006}. With these
data, we cannot differentiate whether this offset is real indicating
the presence of an additional source, or whether it is an
observational artifact caused by the relatively poorer near-infrared
astrometry. The detection of an intermediate dense spectral line
forest with 27 lines within the 4\,GHz bandpass allows the
investigation of additional physical and chemical properties. The 19
CH$_3$OH lines provide temperature estimates toward the two main submm
continuum peaks to $T(\rm{Peak1})\sim 150\pm 50$\,K and
$T(\rm{Peak2})\sim 80\pm 30$\,K, respectively. Searching the dataset
for velocity signatures indicative of rotation around the central
protostar, only a vibrationally-torsionally excited CH$_3$OH line
exhibits a velocity gradient perpendicular to the main molecular
outflow. With a size-scale of a few 1000\,AU and no Keplerian velocity
profile, we are likely not observing a real accretion disk but rather
a larger-scale rotating toroid which may harbor a genuine accretion
disk at its very center.

Setting IRAS\,23151+5912 into context with lower luminosity HMPOs and
higher luminosity UCH{\sc ii}s, its observed physical and chemical
parameters (e.g., line forest, outflow morphology, rotation
signatures) are a mixture of what is observed in the two other
classes. This supports the idea that this region is in a kind of
transition phase between the HMPO and the UCH{\sc ii} region stage.
In a follow-up paper, we will present a comparison of the chemical
properties of this regions with a few sources representative of
different evolutionary stages and luminosities observed all in the
same spectral setup with the SMA (Beuther et al. subm. to A\&A).

\begin{acknowledgements} 
  We like to thank Peter Schilke for providing the XCLASS software to
  model the CH$_3$OH spectra. Furthermore, we appreciated a lot the
  discussions with Thomas Preibisch about the infrared emission of
  this region. In addition, we like to thank the anonymous referee
  whos comments helped to clarify the paper at several points.
  H.B.~acknowledges financial support by the Emmy-Noether-Program of
  the Deutsche Forschungsgemeinschaft (DFG, grant BE2578).
\end{acknowledgements}


\end{document}